\documentclass[%
  reprint,
unsortedaddress,
amsmath,amssymb,
]{revtex4-2}

\usepackage{array}

\usepackage{algorithm}
\usepackage{algpseudocode}

\usepackage{graphicx}% Include figure files
\usepackage{bm}% bold math
\usepackage{epsfig}
\usepackage{epstopdf}

\usepackage{subfigure}

\usepackage{hyperref}% add hypertext capabilities
 \usepackage{tikz}
\usetikzlibrary{arrows.meta, shapes.geometric, positioning, calc, fit, backgrounds}

\definecolor{datacolor}{RGB}{230,245,255}
\definecolor{proccolor}{RGB}{235,255,235}
\definecolor{mathcolor}{RGB}{255,245,230}
\definecolor{outcolor}{RGB}{245,235,255}

\definecolor{theoryblue}{RGB}{230,245,255}
\definecolor{compgreen}{RGB}{235,255,235}
\definecolor{mathorange}{RGB}{255,245,230}

\usepackage{pgfplots}

\usepackage{xcolor}
\usepackage[version=4]{mhchem}

\begin{document}

%\preprint{}

\title{Inversion of Electrochemical Immittance  Spectra  based on the Mellin Transform }

\author{Anis Allagui$^*$}
\email{aallagui@sharjah.ac.ae}

\affiliation{Dept. of Sustainable and Renewable Energy Engineering, University of Sharjah, Sharjah 27272, United Arab Emirates}

\altaffiliation[Also at ]{Center for Advanced Materials Research, Research Institute of Sciences and Engineering, University of Sharjah, Sharjah 27272,  United Arab Emirates}

\affiliation{Dept. of Electrical and Computer Engineering, Florida International University, Miami, FL33174, United States}

\author{Sohaib  Majzoub}
\affiliation{Dept. of Electrical Engineering, University of Sharjah, Sharjah 27272, United Arab Emirates}

\author{Ahmed Elwakil}
\affiliation{Dept. of Electrical Engineering, University of Sharjah, Sharjah 27272, United Arab Emirates}

\affiliation{Nanoelectronics Integrated Systems Center, Nile University, Cairo 12588, Egypt}

\affiliation{Dept. of Electrical and Software Engineering, University of Calgary, Calgary, Alberta T2N 1N4, Canada}

\begin{abstract}

 In this work, we show that the Fredholm integral equations underlying the distribution of relaxation times (DRT), the distribution of capacitive times (DCT), and related frameworks share a common mathematical structure, namely that of a Mellin convolution. This comes from the fact that all standard immittance (impedance or admittance) kernels depend on the product $\omega\tau$ rather than on $\omega$ and $\tau$ independently. Exploiting this structure, we derive an exact algebraic inversion formula in Mellin space that converts the  deconvolution problem into a closed-form relation between the Mellin transform of the measured immittance and that of the unknown distribution function. The framework is validated analytically on a set of   examples including the constant phase element (CPE), the Davidson-Cole (DC) model, and the finite-length Warburg model with blocking boundary conditions. It is also validated numerically using the fast Mellin transform via the fast Fourier transform algorithm for both the CPE and the DC model, including their DRT and DCT recovery under clean and noisy conditions. The approach unifies the impedance- and admittance-based inversions under a single spectral framework,  and provides a new approach  for the   characterization of electrochemical systems from immittance  data.

\end{abstract}

\keywords{Mellin transform, Electrochemical impedance, Tikhonov regularization, Distribution of relaxation times}

\maketitle

\section{Introduction}

Electrochemical impedance spectroscopy (EIS) is among the most widely used characterization techniques in electrochemistry, by providing  frequency-resolved information about charge transfer, diffusion,  interfacial adsorption, as well as electrode geometry\;\cite{huang2020editors, lasia2014electrochemical, vivier2022impedance, Wang:2021aa, doi:10.1021/acsmeasuresciau.2c00070, BOUKAMP2025146892}. The challenge in interpreting EIS data is to identify the physical processes that contribute to the measured response, and at what timescales they operate\;\cite{Orazem_2025}. The standard approach is based on fitting the data with an equivalent circuit model, which requires a topology to be assumed a priori\;\cite{LUKACS2020137199, adfm202109956, HARRINGTON20118005}, and therefore introduces  subjectivity and may limit the generality of the conclusions\;\cite{CIUCCI2019132, MACDONALD20061376}. 
Alternatively, the distribution  of relaxation times (DRT)\;\cite{Orazem_2025, Maradesa:2024aa, WANG2026101789, doi:10.1021/acs.jpcc.5c04766, PLANK2024233845, CHEN20232267, Boukamp_2020}  is based on the mapping of  the impedance in its frequency domain ($Z(j\omega)$, with $\omega$ the angular frequency) to the time domain (i.e. $\tau$-domain of relaxation times) via the Fredholm integral equation of the first kind\;\cite{PLANK2024233845}:
\begin{equation}
Z(j\omega) = \int_0^{\infty} \ker(j\omega, \tau) f(\tau) d\tau
\label{eq:DRT0}
\end{equation}
where $\ker(j\omega, \tau)$ is a kernel function, and  $f(\tau)$  is the desired    function.

As an example, a passive impedance admitting a representation in terms of elementary Voigt   models  (resistor $R$ and capacitor $C$ in parallel, with time constant $\tau=RC$)  can be represented in terms of the discrete  sum\;\cite{garrappa2016models}:
\begin{equation}
Z(j\omega) = R_\infty + \frac{g_1}{1+j\omega\tau_1} + \frac{g_2}{1+j\omega\tau_2} + \ldots
\label{eq:DRTdiscrete}
\end{equation}
where $g_1,g_2\ldots$, $\tau_1,\tau_2\ldots$ are non-negative constants, 
%$\omega$ is the angular frequency, 
 and $R_\infty$ is the high-frequency resistance, 
or more generally    as the continuous superposition of infinite Voigt  processes as\;\cite{Boukamp_2020}:
\begin{equation}
  Z(j\omega) = R_\infty +
  \int_0^\infty \frac{g(\tau)}{1+j\omega\tau}\,d\tau 
  \label{eq:DRT}
\end{equation} 
where $g(\tau)\geq 0$  is the classical  DRT function.
The Voigt  model described by $\ker(j\omega, \tau) = (1+j\omega \tau)^{-1}$ 
corresponds in  the time domain to the response function $\phi(t)$ 
%(obtained by inverse Laplace transform by setting $s=j\omega$)
 as  
%\begin{equation}
$\phi(t) = \mathcal{L}^{-1}\{ k_D (s);t \} =  \tau^{-1} e^{-t/\tau} $,  
%\end{equation}
which has the rate   
%\begin{equation}
 $\Psi(t) = -d\phi(t)/dt= e^{-t/\tau}$\;\cite{garrappa2016models}.  
%\end{equation}

Plotting $g(\tau)$\;vs.\;$\tau$ reveals  the contribution of each relaxation timescale to the total impedance response: sharp peaks  identify well-separated processes, whereas broad features indicate the presence of distributed dynamics\;\cite{BOUKAMP2025146892, Boukamp_2020, batteries5030053, PLANK2024233845}. 
However, recovering $g(\tau)$ from a finite set of noisy or incomplete measurements $\{Z(j\omega_k)\}_{k=1}^N$ is an ill-posed inverse problem that requires some sort of  regularization methods\;\cite{Maradesa:2024aa, WEESE199299, batteries5030053, BOUKAMP201712, PY2025236910, BOUKAMP201535}, which can significantly distort the recovered DRT\;\cite{Boukamp_2020}. Additionally, 
there is a fundamental limitation of the standard DRT model based on the Voigt kernel, which implies   that the impedance tends toward a finite real value at the  limit of very low frequencies\;\cite{PY2024143741}. 
 This does not align well with the physics of systems featuring  blocking electrodes, such as batteries and supercapacitors, for which the impedance   approaches  infinity as the frequency approaches zero due to charge accumulation or diffusion limitations\;\cite{PY2024143741, PhysRevLett.120.116001}. From a mathematical perspective, the DRT is often not well-defined for blocking systems because the imaginary component of the impedance must be integrable for a classical DRT deconvolution, which is not the case for systems exhibiting power-law tails\;\cite{PhysRevLett.120.116001}. 

%Another limitation is that the assumed series connection of RC networks evident from equation (2) is in most cases in-consistent with the actual network structure which might be a fractal  network with embedded self-similar ladder or tree structure.

This led  to the introduction of the distribution of diffusion times (DDT)  by Song and Bazant\;\cite{PhysRevLett.120.116001} 
  to model the admittance rather than impedance  of the system to effectively handle the non-integrable low-frequency behaviors. This translates the problem given by Eq.\;\ref{eq:DRT0} from an  impedance-based DRT framework to an admittance-based formulation\;\cite{PhysRevLett.120.116001}:
\begin{equation}
Y(j\omega)=Z^{-1}(j\omega) = \int_0^{\infty} \ker(j\omega, \tau) p(\tau) d\tau
\label{eq:DRT01}
\end{equation}
where the (diffusion) kernel function can be for instance $\ker(j\omega, \tau) = ( \coth(\sqrt{j\omega \tau})/ \sqrt{j\omega \tau} )^{-1}$ for planar blocking conditions, or $\ker(j\omega, \tau) = ( \tanh(\sqrt{j\omega \tau})/ \sqrt{j\omega \tau} )^{-1}$ for planar transmissive conditions, or others based on the boundary conditions and the symmetry of the system under consideration\;\cite{PhysRevLett.120.116001}. This framework is  considered  mathematically and physically to be more appropriate for describing low-frequency diffusion impedance features\;\cite{PhysRevLett.120.116001}.  
In the same sense,   the distribution of capacitive times (DCT)   was introduced to accommodate  unbounded low-frequency impedances as\;\cite{batteries5030053,PY2024143741}:
\begin{equation}
  Y(j\omega) = G_0 +
  \int_0^\infty \frac{  f(\tau)}{1+ (j\omega\tau)^{-1}}\,d\tau 
  \label{eq:DCT}
\end{equation}
where the kernel is now $\ker(j\omega, \tau) = (1+ (j\omega\tau)^{-1})^{-1}$.   Both the DDP and DCT  approaches allow to map distributions to different  circuit topologies  from the   $RC$ circuit used in the classical DRT structure. 

Other extensions to the EIS inversion problem  include the work of Allagui and Elwakil\;\cite{allagui2024generalized}, wherein the Voigt kernel  in Eq.\;\ref{eq:DRT} is replaced with the Davidson-Cole (DC) model, i.e. $\ker(j\omega, \tau)= (1+(j \omega \tau))^{-\beta}$, or the work of Florsch, Revil, and Camerlynck\;\cite{FLORSCH2014119}, wherein the Havrilliak-Negami (HN) model  is used instead, i.e.  $\ker(j\omega, \tau)= (1+(j \omega \tau)^{\alpha})^{-\beta}$. This implies that if it is known that the  macroscopic system is made of self-similar character at a microscopic level exhibiting a DC or a HN type of response, then the resulting distribution based on these kernels should exhibit lower complexity.

The motivation for the present work is the observation that both Eqs.\;\ref{eq:DRT0} or\;\ref{eq:DRT01} have a precise mathematical structure that has not been previously exploited in the   literature. 
In all   aforementioned cases, the kernels 
$(1+j\omega\tau)^{-1}$,  
$(1+ (j\omega\tau)^{-1})^{-1}$, 
$(1+(j \omega \tau))^{-\beta}$, 
$(1+(j \omega \tau)^{\alpha})^{-\beta}$,
 $\coth(\sqrt{j\omega \tau})/ \sqrt{j\omega \tau}$ 
 and others\;\cite{PhysRevLett.120.116001, Maradesa:2024aa, Huang:2024aa} used for modeling the impedance or admittance (i.e. immittance) depend on the product $\omega\tau$, and the desired distribution function (i.e. DRT $g(\tau)$, DCT $f(\tau)$ or DDT $p(\tau)$) depends on $\tau$ alone.  
 This is the hallmark of a Mellin convolution integral, for which the natural spectral tool is  the Mellin transform.  We show in this study how to convert the immittance inversion  procedure  into an exact algebraic relation in Mellin space (Section\;\ref{sec:MT}). 
 We validate the procedure analytically on a few examples of known impedance and admittance functions, including the constant phase element (CPE) and the Davidson-Cole (DC) model for both DRT and DCT recovery, and the finite-length Warburg   with blocking boundary conditions as a  DDT problem (Section\;\ref{sec:AV}).  
We also present the numerical implementation of the immittance  inversion procedure for the DRT and DCT of both the CPE and the DC model based on the fast Mellin transform (FMT)   using the fast Fourier transform (FFT) algorithm (Section\;\ref{sec:NM}).

\section{Mellin transform formulation}
\label{sec:MT}
  
The Mellin transform of a function $f(t)$,  defined on the positive real axis $0<t<\infty$,  is defined as\;\cite{bateman1954tables, fung1958generalized, bertrand1995mellin}: 
\begin{equation}
\mathcal{M}\{f(t);s\} = \tilde{f}(s) = \int_0^\infty t^{s-1}  f(t) \,dt
\label{eq:Mellin}
\end{equation}
and the inverse is defined as:
\begin{equation}
f(t)= \mathcal{M}^{-1}\{\tilde{f}(s);t\} = \frac{1}{2\pi j } \int_{\sigma- j\infty}^{\sigma +j\infty} \tilde{f}(s) t^{-s} ds
\end{equation}
where $s$ is a complex variable, and the Bromwich contour $\Re(s)=\sigma$ must lie within the fundamental strip of analyticity of $\tilde{f}(s)$. 
The Mellin transform can  be expressed in terms of  the  two-sided Laplace transform by making the change of variable $t=e^{-x}$, which gives\;\cite{bertrand1995mellin}:
\begin{align}
\mathcal{M}\{f(t);s\} 
%= \mathcal{F}\left\{f\!\left(e^{-x}\right);i s \right\} 
= \mathcal{L}\left\{f\!\left(e^{-x}\right);s\right\}
 \end{align}
 The two-sided Laplace transform is defined  as:
\begin{equation}
\mathcal{L}\{g(x);s\} = \int_{-\infty}^{\infty} g(x) e^{-s x} dx
\end{equation}
Additionally, if we write $s= \sigma +j \omega$, we obtain\;\cite{bertrand1995mellin}:
\begin{align}
\mathcal{M}\{f(t); \sigma +j \omega \} 
= \mathcal{F}\left\{ e^{- \sigma x} f\!\left(e^{-x}\right); \xi \right\} 
\label{eq:MTtoFT}
 \end{align}
where the Fourier transform is defined as:
\begin{equation}
\mathcal{F}\{f(x); \xi\} = \int_{-\infty}^{\infty} f(x) e^{-j \xi x} dx
\end{equation}

\begin{figure*}[t]
\centering
\includegraphics[width=\textwidth]{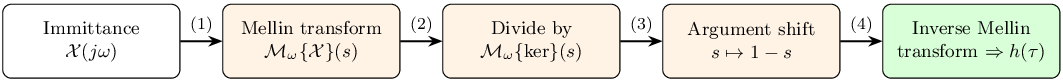}
%\begin{tikzpicture}[
%  node distance=3.725cm,
%  box/.style={
%    rectangle,
%    draw,
%    rounded corners,
%    align=center,
%    minimum width=3cm,
%    minimum height=1.25cm,
%    font=\small
%  },
%  arrow/.style={-{Stealth}, thick},
%  label/.style={font=\footnotesize, midway, above, sloped}
%]
%\node[box, fill=white] (data)
%  {Immittance \\ $\mathcal{X}(j\omega)$};
%\node[box, fill=orange!10, right of=data] (mellin)
%  {Mellin transform  \\ $\mathcal{M}_\omega\{\mathcal{X}\}(s)$};
%\node[box, fill=orange!10, right of=mellin] (divide)
%  {Divide by     \\ $\mathcal{M}_\omega\{\ker\}(s)$};
%\node[box, fill=orange!10, right of=divide] (shift)
%  {Argument shift \\ $s \mapsto 1-s$};
%\node[box, fill=green!15, right of=shift] (output)
%  {Inverse Mellin \\ transform  $\Rightarrow h(\tau)$};
%\draw[arrow] (data) -- 
%  node[label]{(1)} 
%  (mellin);
%\draw[arrow] (mellin) -- 
%  node[label]{(2)} 
%  (divide);
%\draw[arrow] (divide) -- 
%  node[label]{(3)} 
%  (shift);
%\draw[arrow] (shift) -- 
%  node[label]{(4)} 
%  (output);
%\end{tikzpicture}
\caption{Flowchart of the Mellin-based immittance inversion procedure. The measured immittance $\mathcal{X}(j\omega)$ (impedance $Z(j\omega)$ or admittance $Y(j\omega)$) is mapped to its 
corresponding distribution function $h(\tau)$ via four steps in Mellin space. The kernel $\ker(j\omega,\tau)$ is chosen 
according to the framework of interest: Voigt for DRT, capacitive for DCT, or diffusion-based for DDT.}
\label{fig:flowchart}
\end{figure*}

The Mellin transform has many interesting properties that can be found in Erd\'{e}lyi et al.\;\cite{bateman1954tables}, but the  key property for the present work is  the Mellin (generalized or multiplicative\;\cite{fung1958generalized}) convolution theorem. For an integral of the form\;\cite{bateman1954tables}: 
\begin{equation}
h(t) = t^a \int_0^\infty \xi^{\beta} f_1\!\left( t\, \xi \right) f_2( \xi )\,{d\xi}
\end{equation}
the Mellin transform is given by: 
\begin{equation}
\tilde{h}(s) = 
\tilde{f_1}(s+a)
\tilde{f_2}(1-s-a+\beta)
\end{equation} 
which converts the Mellin convolutions into point-wise products in the $s$ domain,  exactly as the Fourier convolution theorem converts ordinary convolutions into products in the frequency domain. 
Comparing with Eqs.\;\ref{eq:DRT0} or\;\ref{eq:DRT01}, the Mellin transform of an immittance function $\mathcal{X}_r(j\omega)$ is given by:
\begin{equation}
\mathcal{M}_{\omega}\{\mathcal{X}_r(j\omega);s\}
= \mathcal{M}_{\omega}\{\ker (j\omega);s\}
 \mathcal{M}_{\tau}\{g (\tau);1-s\}
 \label{eqMTX}
\end{equation}
from which we obtained the main result of this study:
\begin{equation}
\mathcal{M}_{\tau}\{g (\tau);s\} = \frac{\mathcal{M}_{\omega}\{\mathcal{X}_r(j\omega);1-s\}} {\mathcal{M}_{\omega}\{\ker (j\omega);1-s\}} 
\label{eqMTg}
\end{equation}
after substituting $s \mapsto 1-s$. 
We remark the analogy that  a transfer function of a linear system is obtained as the ratio of output to input spectra in Fourier space. Here the Mellin transform of the distribution function $g(\tau)$ is obtained as the ratio of the immittance spectrum to the kernel spectrum in Mellin space. 
The distribution $g(\tau)$ in the $\tau$ domain is then recovered by Mellin inversion of Eq.\;\ref{eqMTg}. Fig.\;\ref{fig:flowchart} provides   a flowchart summary of the proposed Mellin-based immittance inversion procedure.

We note that  the exact inversion formula in Eq.\;\ref{eqMTg}  require no regularization at the analytical level.  
In practical implementations with numerical data, however, the Mellin  integrals are evaluated over a finite frequency window,  and usually  in the presence of noise. This necessitates spectral truncation and/or windowing in Mellin space, which act as   implicit  regularization themselves\;\cite{bertrand1995mellin, IRINO2002181, 845927, De-Sena:2007aa} (see Section\;\ref{sec:NM}).

\section{Analytical examples}
\label{sec:AV}

\subsection{Distribution of relaxation times}

As a first example to validate the  inversion procedure, we consider the reduced impedance function defined from Eq.\;\ref{eq:DRT} with the removal of the $R_\infty$ offset in the DRT sense as:
\begin{equation}
 \mathcal{X}_r(j\omega)= Z_r(j\omega) = Z(j\omega)-R_\infty = 
  \int_0^\infty \frac{g(\tau)}{1+j\omega\tau}\,d\tau
  \label{eq:Zr}
\end{equation}
This is clearly   a Mellin convolution of the DRT function $g(\tau)$ with the   basis function $(1+j\omega\tau)^{-1}$ (Voigt model).  
Taking the Mellin transform of $Z_r(j\omega)$ 
 with respect to the variable $\omega$ gives:
\begin{equation}
\mathcal{M}_{\omega}\{Z_r(j\omega);s\}
= \pi (j)^{-s}\csc(\pi s)\,
\mathcal{M}_\tau\left\{ g(\tau) ;1-s\right\}
\label{eq:main}
\end{equation}
where we used the result\;\cite{bateman1954tables}:
\begin{equation}
\mathcal{M}\{ (1+a t)^{-\nu} ;s\} = a^{-s} B(s,\nu-s),\; 0<\text{Re}(s)<\text{Re}(\nu)
\end{equation}
 For $\nu=1$, the Mellin transform of the Voigt  kernel with respect  $\omega$ simplifies to:
\begin{equation}
  \mathcal{M}_\omega\left\{ ({1+j\omega\tau})^{-1};\,s\right\}
  = \pi (j\tau)^{-s} \csc(\pi s),\; 0<\text{Re}(s)<1 
  \label{eq:MTDebye}
\end{equation}
The strip $0<\Re(s)<1$ is the fundamental strip of the Voigt kernel's Mellin transform, and therefore  the inversion contour $\Re(s)=\sigma$ must
lie within this strip. 
Note that the Mellin argument shift from $s$ to $1-s$ in Eq.\;\ref{eq:main} absorbs the term $\tau^{-s}$ that comes from Eq.\;\ref{eq:MTDebye}.   
Solving for the DRT, we obtain:
\begin{equation} 
\mathcal{M}_\tau\left\{ g(\tau) ;1-s\right\} 
 = \pi^{-1}  \sin(\pi s) (j)^{s}  \mathcal{M}_{\omega}\{Z_r(j\omega);s\} 
\label{eq:inverse}
\end{equation}
and then  with  $1-s \mapsto s$ we obtain:
\begin{equation} 
\mathcal{M}_\tau\left\{ g(\tau) ;s\right\} 
%\tilde{g}(s) 
= \pi^{-1}  \sin(\pi (1-s)) (j)^{1-s}  \mathcal{M}_{\omega}\{Z_r(j\omega);1-s\} 
\label{eq:inverse2}
\end{equation}
The DRT function $g(\tau)$ is recovered from by Mellin inversion (Fig.\;\ref{fig:flowchart}). The relations given by Eqs.\;\ref{eq:main}, \ref{eq:MTDebye} and thus Eq.\;\ref{eq:inverse2}  are exact, and holds for any admissible impedance function.

We apply the procedure above to the case of the CPE, whose impedance is defined as\;\cite{allagui2024procedure}: 
\begin{equation}
Z_c(j\omega) =  {R_0}{(j\omega\tau_c)^{-\alpha}} = \int_0^\infty \frac{g_c(\tau)}{1+j\omega\tau}\,d\tau 
\label{eq:CPE}
\end{equation}
where $0 < \alpha < 1$, $R_0$ is the resistance, and $\tau_c$ is a characteristic time constant.
Its Mellin transform  is:
\begin{equation}
\mathcal{M}_\omega\{Z_c(j\omega); s\} = 2\pi R_0 (j\tau_c)^{-\alpha}    \, \delta(s - \alpha)
\label{eq:MZ}
\end{equation}
and therefore
\begin{equation}
\mathcal{M}_\tau\{g(\tau); (1-s) \} = 2 R_0 \tau_c^{-\alpha} j^{s-\alpha} \sin(\pi s) \, \delta(s - \alpha)
\label{eq:Mg1}
\end{equation}
which we then rewrite as:
\begin{equation}
\mathcal{M}_\tau\{g(\tau);s\} = 2 R_0 \tau_c^{-\alpha} j^{1-\alpha -s} \sin(\pi s) \, \delta(s-(1 - \alpha))
\label{eq:Mg1}
\end{equation}
noting that $\sin(\pi(1-s))=\sin(\pi s)$.  The inverse Mellin transform with  $s=c+j\xi$ for $\xi \in ]-\infty;\infty[$ ($ds=j\,d\xi$), and $c=1-\alpha$, is:
\begin{align}
g_c(\tau)
&= \frac{1}{2\pi j} \int_{c-j\infty}^{c+j\infty}
   2 R_0\, \tau_c^{-\alpha}\, j^{1-\alpha-s}\, \sin(\pi s)\,
   \delta(s-(1-\alpha))\, \tau^{-s}\, ds 
   \notag \\
&= \frac{1}{2\pi} \int_{-\infty}^{+\infty}
   2 R_0\, \tau_c^{-\alpha}\, j^{1-\alpha-s}\, \sin(\pi s)\,
   \delta(\xi)\, \tau^{-s}\, d\xi
\end{align}
where we used $\delta(s-(1-\alpha)) = \delta(j\xi) = \delta(\xi)/|j| = \delta(\xi)$. With the use of the translation property   of the Dirac delta function $\int_{-\infty}^{\infty} f(t) \delta(t-T) dt = f(T)$, i.e. with the evaluation at $s=1-\alpha$,  
%\begin{equation}
%\int_{-\infty}^{\infty} f(t) \delta(t-T) dt = f(T)
%\end{equation}
we obtain the well-known  power law in $\tau$\;\cite{allagui2024generalized, allagui2024procedure}
\begin{equation}
 {g_c(\tau) = \pi^{-1} R_0\, \sin(\pi\alpha)\; \tau_c^{-\alpha}\, \tau^{\alpha-1}}
\label{eq:gCPE}
\end{equation} 
Note that $g_c(\tau) > 0$ for all $\tau > 0$, which is 
consistent with the physical requirement that the DRT be non-negative. The integral $\int_0^\infty g_c(\tau)\,d\tau$ 
diverges, which is in line with the absence of a 
characteristic relaxation timescale in the CPE.

We apply   the same procedure again  to the case of  the DC impedance   given by\;\cite{davidson1951dielectric}:
\begin{equation}
Z_{{d}}(j\omega)= \frac{R_0}{(1+ j \tau_0 \omega)^{\alpha}}
\label{eq:Zd}
\end{equation}
Its Mellin transform with respect to $\omega$ is\;\cite{bateman1954tables}:
\begin{equation}
\mathcal{M}_{\omega}\{Z_d(j\omega);s\} 
= R_0 \frac{  \Gamma(s) \Gamma(\alpha-s)}{  (j \tau_0)^{s} \Gamma(\alpha) }
%= R_0 (j \tau_0)^{-s} B(s,\alpha-s)
\end{equation}
which is analytic for $0<\Re(s)<\alpha$ (this is the fundamental strip of the DC impedance's Mellin transform),  
and therefore $\mathcal{M}_\tau\left\{ g_d(\tau) ;s\right\}$ is obtained after substituting $s \mapsto 1-s$ as:
\begin{equation} 
\mathcal{M}_\tau\left\{ g_d(\tau) ;s\right\} 
%\tilde{g}(s) 
= R_0  (\tau_0)^{s-1}   \frac{  \Gamma(s+\alpha-1)  }{  \Gamma(\alpha) \Gamma(s)   } 
\label{eq:inverse20}
\end{equation}
Applying the inverse Mellin transform with respect to $\tau$ gives the expected result\;\cite{allagui2024generalized, allagui2024procedure}:
\begin{align}
g_d(\tau) &= \frac{R_0 \sin (\alpha \pi   )}{ \pi    \tau   (\tau_{0}/\tau -1)^{\alpha }}, \;\; \tau< \tau_{0} \nonumber \\
& = 0, \;\; \text{otherwise}
\label{eq:gd}
\end{align}
We verify that in this case:
\begin{equation}
\int_0^{\infty} g_d(\tau) d\tau = R_0
\end{equation}

\subsection{Distribution of capacitive times}

We now apply the DCT framework to the CPE admittance, which is defined as: 
\begin{equation}
Y_c(j\omega) = R_0^{-1} {(j\omega\tau_c)^\alpha} = 
\int_0^\infty \frac{f_c(\tau)}{1+(j\omega\tau)^{-1}}\,d\tau
\label{eq:CPE_DCT}
\end{equation}
Its Mellin transform is:
\begin{equation}
\mathcal{M}_\omega\{Y_c(j\omega);\,s\} 
= 2\pi R_0^{-1}(j\tau_c)^\alpha\,\delta(s+\alpha)
\label{eq:MY_CPE}
\end{equation}
and that of the DCT kernel is:
\begin{equation}
\mathcal{M}_\omega\left\{(1+(j\omega\tau)^{-1})^{-1};\,s\right\}
= -\pi(j\tau)^{-s}\csc(\pi s)
\end{equation}
with $-1<\Re(s)<0$, so the inversion contour for the DCT must satisfy $-1<\Re(s)<0$. 
Then
 \begin{equation}
\mathcal{M}_\tau\{f_c(\tau)\}(1-s) 
=   -2R_0^{-1}(j\tau_c)^\alpha(j)^{s}\sin(\pi s)\,\delta(s+\alpha)
\label{eq:Mf_CPE}
\end{equation}
Evaluating the delta function at $s=-\alpha$ gives:
\begin{equation}
\mathcal{M}_\tau\{f_c(\tau)\}(1-s) 
= 2R_0^{-1}\tau_c^\alpha\sin(\pi\alpha)\,\delta(s+\alpha)
\end{equation}
Applying the argument shift $s\mapsto 1-s$:
\begin{equation}
\mathcal{M}_\tau\{f_c(\tau)\}(s) 
= 2R_0^{-1}\tau_c^\alpha\sin(\pi\alpha)\,\delta(s-(1+\alpha))
\label{eq:Mf_CPE2}
\end{equation}
Applying the inverse Mellin transform on the Bromwich contour 
$c=1+\alpha$, with $s=(1+\alpha)+j\xi$ and $ds=j\,d\xi$:
\begin{align}
f_c(\tau) 
&= \frac{1}{2\pi j}\int_{c-j\infty}^{c+j\infty}
2R_0^{-1}\tau_c^\alpha\sin(\pi\alpha)\,
\delta(s-(1+\alpha))\,\tau^{-s}\,ds \notag \\
&= \frac{1}{2\pi}\int_{-\infty}^{+\infty}
2R_0^{-1}\tau_c^\alpha\sin(\pi\alpha)\,
\delta(\xi)\,\tau^{-s}\,d\xi
\end{align}
Applying the sifting property of $\delta(\xi)$ at $\xi=0$, 
i.e.\;$s=1+\alpha$, results in :
\begin{equation}
f_c(\tau) = 
 { \pi^{-1} R_0^{-1} \sin(\pi\alpha)}\,\tau_c^\alpha\,\tau^{-(1+\alpha)}
\label{eq:fCPE}
\end{equation}
which is also a pure power law in $\tau$. 
 As in the DRT case, the integral $\int_0^\infty f_c(\tau)\,d\tau$ also 
diverges.

As a second example, we consider the reduced admittance given by the inverse of the DC impedance in Eq\;\ref{eq:Zd}, i.e.:
\begin{equation}
  Z_d^{-1}(j\omega) = {R_0^{-1}}{(1+ j \tau_0 \omega)^{\alpha}} = 
  \int_0^\infty \frac{  f_d(\tau)}{1+ (j\omega\tau)^{-1}}\,d\tau 
  \label{eq:DCT}
\end{equation}
With:
 \begin{equation}
  \mathcal{M}_\omega\left\{ ({1+(j\omega\tau)^{-1}})^{-1};\,s\right\}
  = -\pi (j\tau)^{-s} \csc(\pi s) 
\end{equation}
 and 
\begin{equation}
\mathcal{M}_{\omega}\{Z_d^{-1}(j\omega);s\} 
=  \frac{ \Gamma(s) \Gamma(-\alpha-s)}{ R_0 (j \tau_0)^{s} \Gamma(-\alpha) }
\end{equation}
we obtain the corresponding Mellin-transformed DCT as:
\begin{equation} 
\mathcal{M}_\tau\left\{ f_d(\tau) ;s\right\} 
= -R_0^{-1}  (\tau_0)^{s-1}   \frac{  \Gamma(s-\alpha-1)  }{  \Gamma(-\alpha) \Gamma(s)   } 
\label{eq:inverse202}
\end{equation}
The inverse Mellin transform with respect to $\tau$ gives:
\begin{align}
f_d(\tau) &= \frac{ (\tau_{0}/\tau -1)^{\alpha } \sin (\alpha \pi   )}{ \pi R_0     \tau   }, \;\; \tau< \tau_{0} \nonumber \\
& = 0, \;\; \text{otherwise}
\label{eq:fd}
\end{align} 
Unlike the DRT case, the integral $\int_0^{\infty} f_d(\tau) d\tau$ diverges, which is consistent with the unbounded low-frequency behavior of the DC admittance.

\subsection{Distribution of diffusion times}

We consider here the simple case of finite-length Warburg admittance of a planar electrode system with 
blocking boundary conditions, defined as\;\cite{PhysRevLett.120.116001}:
\begin{align}
Y_{w}(j\omega) 
&= \left({R_0} \frac{\coth(\sqrt{j\omega\tau_0})}{\sqrt{j\omega\tau_0}}\right)^{-1}
\nonumber \\
&= \int_0^\infty 
  \left( \frac{\coth(\sqrt{j\omega\tau})}{\sqrt{j\omega\tau}}\right)^{-1}\,p(\tau)\,d\tau
\label{eq:Yflw}
\end{align}
where  the DDT function $p(\tau)$ is to be determined\;\cite{PhysRevLett.120.116001}. This  problem  represents a single blocking diffusion element, which should give immediately $p(\tau) =  {R_0^{-1}}\,\delta(\tau-\tau_0)$ (sifting  property of the delta function), that we verify below based on our theory.

First,  for the Mellin transform of the kernel given by $\ker_w(j\omega,\tau)= \sqrt{j\omega\tau} \tanh(\sqrt{j\omega\tau})$, we have after the substitutions $u = \sqrt{j\omega\tau}$,  
$\omega = u^2/(j\tau)$, $d\omega = 2u\,du/(j\tau)$:
\begin{equation}
\mathcal{M}_\omega\left\{ \ker_w(j\omega,\tau) ;\,s \right\}
=  \frac{2}{(j\tau)^{s}} \int_0^\infty u^{2s} \tanh(u) du
\label{eq:kernel_MT_tanh}
\end{equation}
We now substitute with 
the series representation of $\tanh(z)$ given by\;\cite{tanh}
\begin{equation}
\tanh(z) = 8 z \sum\limits_{k=1}^{\infty} \frac{1}{\pi^2 (2k-1)^2+ 4 z^2},\; \frac{i z}{\pi}-\frac{1}{2} \notin \mathbb{Z}
\label{eq:tanh}
\end{equation}
  to obtain:
\begin{align}
\int_0^\infty u^{2s} & \tanh(u)\,du 
= \sum_{k=1}^\infty
\int_0^\infty
\frac{2 u^{2s+1}}{u^2+\left(k- {1}/{2}\right)^2\pi^2}
\,du 
\\ & = -\sum_{k=1}^\infty  \pi^{2s+1}  (k-1/2)^{2s} \csc (\pi s) 
\\ & = -\pi^{2s+1} \csc (\pi s) \sum_{k=1}^\infty (k-1/2)^{2s}
\\ & = \pi^{2s+1} \csc (\pi s) (1-4^{-s}) \zeta(-2s)
\label{eq:tanh_split}
\end{align}
where  $\zeta(z) = \sum_{k=1}^{\infty} k^{-z}$ is the Riemann zeta function. 
Substituting back into Eq.\;\ref{eq:kernel_MT_tanh} gives:
\begin{equation}
\mathcal{M}_\omega\left\{ \ker_w(j\omega,\tau) ;\,s \right\}
= 
\frac{2 \pi^{2s+1}  (1-4^{-s}) \zeta(-2s) }{(j\tau)^s  \sin (\pi s) }  
\label{eq:kernel_MT_tanh_final}
\end{equation}
It is worth noting that this result is not a classical 
Mellin transform in the strict sense, since the kernel 
$\ker_w(j\omega,\tau) = \sqrt{j\omega\tau}\tanh(\sqrt{j\omega\tau})$ 
grows as $\sqrt{j\omega\tau}$ as $\omega\to\infty$, and 
therefore the defining integral diverges at the upper limit 
for any $\text{Re}(s)>0$. The result is thus understood 
in the sense of analytic continuation.

Second, since $Y_{w}(j\omega) = R_0^{-1}\ker_w(j\omega,\tau_0)$, 
we replace $\tau\to\tau_0$ in 
Eq.\;\ref{eq:kernel_MT_tanh_final} to obtain:
\begin{equation}
\mathcal{M}_\omega\{Y_{w}(j\omega);\,s\} 
= 
\frac{2 \pi^{2s+1}  (1-4^{-s}) \zeta(-2s) }{R_0(j\tau_0)^s  \sin (\pi s) } 
\label{eq:MY_flw}
\end{equation}
Now solving for  $\mathcal{M}_\tau\{p\,\}(1-s)$ by making sure the term $\tau^{-s}$ is properly accounted for when the argument is shifted from $s$ to $1-s$ as done above, gives  $\mathcal{M}_\tau\{p(\tau);1-s\} =  {\tau_0^{-s}}/{R_0}$, 
and thus
\begin{equation}
\mathcal{M}_\tau\{p(\tau);s\} = {R_0}^{-1} {\tau_0^{s-1}}
\end{equation}
	The inverse Mellin transform  gives, as expected, a delta function concentrated at $\tau_0$:
\begin{equation}
p(\tau) =  {R_0^{-1}}\,\delta(\tau-\tau_0)
\label{eq:f_flw}
\end{equation} 
%We verify by substituting  Eq.\;\ref{eq:f_flw} back into  Eq.\;\ref{eq:Yflw} and applying the sifting property of the delta function gives immediately $Y_{w}(j\omega)$.  
The normalization is also verified from $\int_0^\infty p(\tau)\,d\tau 
= {R_0^{-1}}$. 
Eq.\;\ref{eq:f_flw} means again that the system is characterized by a single diffusion time scale $\tau_0=L^2/D$ ($L$ is the diffusion 
length and $D$ the diffusion coefficient).  This is the     analog of the DRT of a simple $RC$ circuit, which is also a single delta function concentrated at the time constant $\tau_0 = RC$.

\section{Numerical Implementation}
\label{sec:NM}

In this section, we present  examples of numerical implementation of the Mellin transform-based  procedure of Fig.\;\ref{fig:flowchart} for both   the CPE and DC model. We describe the Fast Mellin Transform (FMT) implementation on a logarithmic frequency grid   using the standard FFT algorithm\;\cite{De-Sena:2007aa}. 

\subsection{Mellin–FFT Inversion Algorithm}

 The whole numerical procedure is summarized in Algorithm\;\ref{alg}.  
With the change of variable $\omega = e^{-x}$ and with $s = \sigma + j\xi$, we can write the Mellin transform of the immittance function $\mathcal{X}_r(j\omega)$ in Eq.\;\ref{eqMTX} in a Fourier-transform form as\;\cite{IRINO2002181}: 
\begin{equation}
\mathcal{M}_\omega\{\mathcal{X}_r(j\omega); \sigma + j\xi\} 
= \mathcal{F}\{ e^{-\sigma x} \mathcal{X}_r(j e^{-x}); \xi \}
\end{equation} 
Numerically, this consists of first discretizing  the $x$-axis, such that $x_n = -L + n\,\Delta x$ for $n = 0,\ldots,N-1$ ($\Delta x = 2L/N$). 
%We took $N=2^{16}$ and  $L=30$. 
The sampled angular frequencies are therefore $\omega_n = e^{-x_n}$ (Steps 1 and 2 in Algorithm\;\ref{alg}). 
 The forward  Mellin transform is then computed by FFT through:
 \begin{equation}
 \tilde{X}(\sigma + j \xi_k)
  = \Delta x\; e^{- {j}\xi_k x_0}\;
    \texttt{FFT}\!\left\{e^{-\sigma x_n} X( {j} e^{-x_n})\right\}_k
\end{equation}
with $x_0=-L$ and $\xi_k$ are the discrete Fourier frequency bins ($\xi_k = k/(N\,\Delta x)$, $k=0,1,\ldots,N/2-1,-N/2,\ldots,-1$) (Step 3).  The phase factor $e^{- {j}\xi_k x_0}$ is essential as it  compensates for the fact
that the FFT assumes samples starting at $x= 0$, while the actual grid starts at $x=-L$. This is then followed by the argument shift   $s \mapsto 1-s$, which   is implemented by an index reflection
in the Mellin-frequency array (Step 4). 

Next, 
the   division given by Eq.\;\ref{eqMTg}    is numerically ill-posed  and requires stabilization. Because the kernel is conjugate-symmetric on the contour,   we therefore restrict the division to the better-conditioned half to avoid having unstable noise from the high $|\xi|$ tail. 
We also adopt a Tikhonov-regularized division (with   parameter $\lambda$) to obtain: 
\begin{equation}
\tilde{h}_{\lambda}(s) =  \frac{ \tilde{\mathcal{X}}_r(1-s)  \tilde{\ker}(1-s)^* }{ |\tilde{\ker}(1-s)|^2 + \lambda }
 \end{equation}
 which is then multiplied by a Hann window:
 \begin{equation}
W(\xi)= \frac{1}{2} \left[ 1 + \cos\left( \frac{\pi \xi}{\xi_{\text{cut}}} \right) \right]
\end{equation}
 to suppress Gibbs type oscillations. Together they act as a low-pass denoising filter (Step 6). 
 
 Then, because the desired distribution function is real-values, its Mellin-domain transform satisfies the conjugate symmetry. This is what is implemented in Algorithm\;\ref{alg} under Step 7, Hermitian completion. 
 
 After assembling the stabilized Mellin-domain spectrum, the inverse Mellin transform back to the $\tau$-domain at $\tau = e^{-y}$ is obtained via: 
\begin{equation}
h_n =   e^{\sigma_h x_n}\,
    \operatorname{Re}\!\left[\Delta x^{-1}\,
    \texttt{IFFT}\!\left\{\tilde{h}(\xi_k)\, e^{ {j}\xi_k x_0}\right\}_n\right]
\end{equation}
The phase factor $e^{j \xi_k x_0}$ is the inverse counterpart of the phase factor used in the forward transform.

A critical step for successful inversion
is the choice of the Mellin abscissa $\sigma_h$ for the inverse
transform. This must lie within the fundamental strip of the
distribution function's Mellin transform. For the CPE DRT, the
relevant Mellin pole is at $s=\alpha$, so the natural choice is
$\sigma_h = 1-\alpha$ (we use $\sigma_h=0.25$ for $\alpha=0.75$). For
the CPE DCT, the pole is at $s=-\alpha$, giving $\sigma_h = 1+\alpha$
(we use $\sigma_h=1.75$). For the DC model, whose DRT has a strip of
analyticity $0<\Re(s)<\alpha$, the centre of the strip gives
$\sigma_h = 1-\alpha/2$ (we use $\sigma_h=0.75$ for $\alpha=0.5$). In
general, $\sigma_h$ can be determined adaptively from the high- and
low-frequency power-law exponents of the measured immittance, without
prior knowledge of the underlying physical model.

\begin{algorithm}[H]
\caption{Mellin--FFT inversion}
\label{alg}
\begin{algorithmic}[1]
\Require Immittance data $\{(\omega_n, X(\mathrm{j}\omega_n))\}_{n=0}^{N-1}$,
         regularisation parameter $\lambda$, cutoff $\xi_{\mathrm{cut}}$,
         Mellin abscissa $\rho$, half-width $L$, step $\Delta x$
%\Ensure  $\{(e^{-x_n},\, h_n)\}_{n=0}^{N-1}$

\State \textbf{Step 1: Build grids}
\State Construct log-frequency grid:
\[
  x_n = -L + n\,\Delta x, \quad n = 0,\ldots,N-1
\]
\State Construct Mellin-frequency grid:
\[
  \{\xi_k\} = 2\pi\,\texttt{fftfreq}(N,\Delta x)
\]

\vspace{0.5em}
\State \textbf{Step 2: Evaluate immittance on the grid}
\State Set $\omega_n = e^{-x_n}$ and evaluate immittance at each grid point

\vspace{0.5em}
\State \textbf{Step 3: Forward FMT   at abscissa $\sigma = \rho$}
%\State Form the weighted sequence $f_n = e^{-\sigma x_n} X(\mathrm{j}\omega_n)$
\State Compute FFT and apply phase correction and scaling:
\[
  \tilde{X}_k
  = \Delta x\; e^{- {j}\xi_k x_0}\;
    \texttt{FFT}\!\left\{e^{-\sigma x_n} X( {j}\omega_n)\right\}_k
\]

\vspace{0.5em}
\State \textbf{Step 4: Argument reflection}
% to the complementary contour}
\State Map spectrum to the reflected contour $s \mapsto 1-s$:
\[
  \tilde{X}^{(1-s)}_k = \tilde{X}_{-k \bmod N}
\]

\vspace{0.5em}
\State \textbf{Step 5: Evaluate Mellin kernel on the reflected contour}
\State Set $q_k = \sigma -  {j}\xi_k$ and compute the kernel transform:
\[
  \tilde{K}_k
  = \frac{\pi\, e^{- {j}\pi q_k/2}}{\sin(\pi q_k)}
\]
(use the Voigt kernel as written, or its negative for the capacitive kernel)

\vspace{0.5em}
\State \textbf{Step 6: Stabilized algebraic inversion on the stable half $\xi_k \in [-\xi_{\mathrm{cut}},\,0]$}
\State Apply Tikhonov regularization with Hann windowing:
\[
  \tilde{h}_k
  = \tilde{X}^{(1-s)}_k\,\tilde{K}_k^*
    \!\left(|\tilde{K}_k|^2 + \lambda\right)^{-1}
    W_k,
  \;\;
  W_k = \tfrac{1}{2}\!\left[1 + \cos\!\left(\pi\xi_k/\xi_{\mathrm{cut}}\right)\right]
\]

\vspace{0.5em}
\State \textbf{Step 7: Hermitian completion}
\State For every $k$ on the stable side set $\tilde{h}_{-k \bmod N} \leftarrow \tilde{h}_k^*$;
       force $\tilde{h}_0 \in \mathbb{R}$

\vspace{0.5em}
\State \textbf{Step 8: Inverse FMT at abscissa $\sigma_h = 1-\rho$}
\State Apply phase correction and compute IFFT, then rescale:
\[
  h_n
  = e^{\sigma_h x_n}\,
    \operatorname{Re}\!\left[\Delta x^{-1}\,
    \texttt{IFFT}\!\left\{\tilde{h}_k\, e^{ {j}\xi_k x_0}\right\}_n\right]
\]

\vspace{0.5em}
\State \textbf{Step 9: Optional positivity clip}
\State $h_n \leftarrow \max(h_n,\, 0)$

\State \Return $\left\{\!\left(e^{-x_n},\, h_n\right)\!\right\}_{n=0}^{N-1}$
\end{algorithmic}
\end{algorithm}

\subsection{CPE model DRT and DCT recovery}

\begin{figure*}[t]
  \centering
  \includegraphics[width=\textwidth]{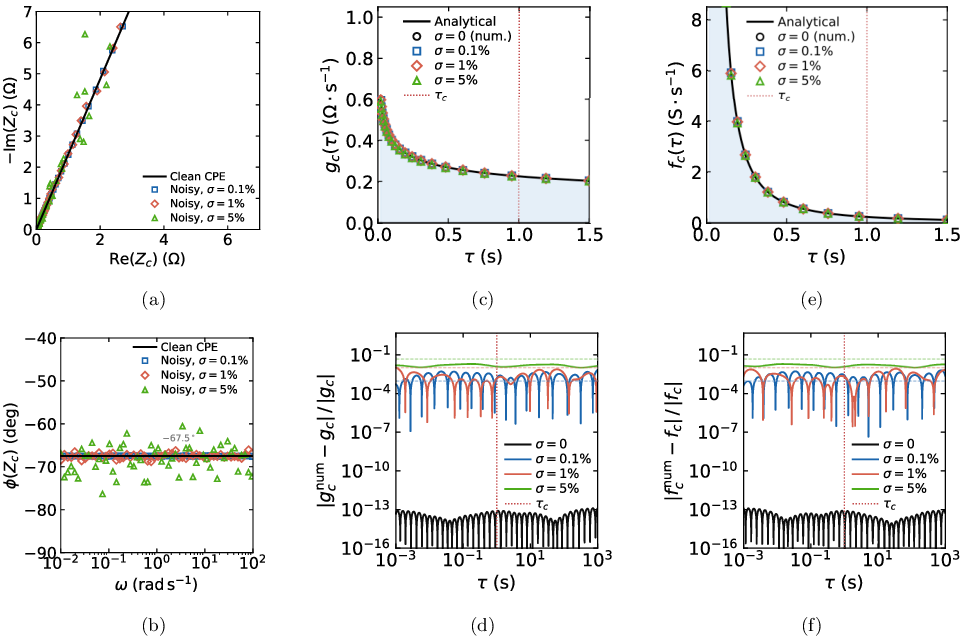}
%  \includegraphics[height=0.285\linewidth]{figures/panel_a_nyquist.pdf}\hfill
%  \includegraphics[height=0.285\linewidth]{figures/panel_c_drt.pdf}\hfill
%  \includegraphics[height=0.285\linewidth]{figures/panel_e_dct.pdf}\\[2pt]
%  \makebox[0.325\linewidth]{\small(a)}\hfill
%  \makebox[0.325\linewidth]{\small(c)}\hfill
%  \makebox[0.325\linewidth]{\small(e)}\\[6pt]
%  \includegraphics[height=0.285\linewidth]{figures/panel_b_bode.pdf}\hfill
%  \includegraphics[height=0.285\linewidth]{figures/panel_d_drt_err.pdf}\hfill
%  \includegraphics[height=0.285\linewidth]{figures/panel_f_dct_err.pdf}\\[2pt]
%  \makebox[0.325\linewidth]{\small(b)}\hfill
%  \makebox[0.325\linewidth]{\small(d)}\hfill
%  \makebox[0.325\linewidth]{\small(f)}
  \caption{
  Numerical results for the CPE (of impedance $Z_c(j\omega) =  {R_0}{(j\omega\tau_c)^{-\alpha}}$) with $R_0=1\,\Omega$,
  $\tau_c=1\,\mathrm{s}$, $\alpha=0.75$, and noise levels
  $\sigma\in\{0,0.1\%,1\%,5\%\}$.
  \textbf{(a)}~Nyquist plot of $Z_c(j\omega)$: the clean CPE (solid black) traces a
  straightline at $-67.5^\circ$;  markers show the noisy data.
  \textbf{(b)}~corresponding Bode phase vs.\ $\omega$ (log scale).
  \textbf{(c)}~DRT $g_c(\tau)$ recovery on linear axes: solid black
  curve is the analytical Eq.~\eqref{eq:gCPE};   markers are the
  Mellin-FFT recovery including the noise-free case ($\sigma=0$). 
  Red dotted line marks $\tau_c=1\,\mathrm{s}$.
  \textbf{(d)}~Pointwise relative error of the DRT recovery; dashed
  horizontal lines mark the input noise level $\sigma$. The $\sigma=0$
  curve stays near $10^{-14}$.
  \textbf{(e)}~DCT $f_c(\tau)$ recovery; same conventions as~(c).
  \textbf{(f)}~Pointwise relative error of the DCT recovery; same
  conventions as~(d).}
  \label{fig:cpe_results}
\end{figure*}

Fig.\;\ref{fig:cpe_results} shows the Nyquist plot of
$Z_c(j \omega)=R_0(  j \omega\tau_c)^{-\alpha}$ for $R_0=1\,\Omega$,
$\tau_c=1\,\mathrm{s}$, $\alpha=0.75$ (clean CPE), as well as noisy
data generated using~\cite{SACCOCCIO2014470}:
\begin{equation}
  Z_m(  j \omega_k)
    = Z_c(  j \omega_k)\left[1+\sigma\varepsilon_{R,k}
      +  j \sigma\varepsilon_{I,k}\right] 
  \label{eq:noise-Z}
\end{equation}
where $\varepsilon_{R,k},\varepsilon_{I,k}\sim\mathcal{N}(0,1)$, and
$\sigma\in\{0,0.1\%,1\%,5\%\}$. Fig.\;\ref{fig:cpe_results}(a) shows the Nyquist plot, where  the clean CPE traces an inclined straight line at $-67.5^\circ$. This can be clearly seen from the Bode phase plot in Fig.\;\ref{fig:cpe_results}(b). The recovered DRT and DCT are shown
in Fig.\;\ref{fig:cpe_results}(c) and Fig.\;\ref{fig:cpe_results}(e) for all noise levels, with  the corresponding errors   in
Figs.\;\ref{fig:cpe_results}(d) and\;\ref{fig:cpe_results}(f).

For the noise-free case ($\sigma=0$), we set $N=2^{16}$, $L=30$,
$\sigma_{h,\mathrm{DRT}}=0.25$, $\sigma_{h,\mathrm{DCT}}=1.75$, and
$\{\xi_\mathrm{cut},\lambda\}=\{10,10^{-20}\}$. The recovered
distribution attains a relative error of order $10^{-13}$--$10^{-14}$
across twelve decades in $\tau$, which is essentially the
floating-point limit for double-precision FFT arithmetic with
$N=2^{16}$ samples.  We note that the Tikhonov parameter $\lambda$
and cutoff $\xi_\mathrm{cut}$ for the noisy cases were determined by
minimising the $L^2$ relative error against the analytical reference
distributions. For experimental data lacking an analytical reference,
these can be tuned using standard regularisation selection criteria
such as the L-curve or discrepancy principle.

For the noisy data, we kept the same general parameters of the
algorithm, but adjusted the stabilization parameters:
$\{\sigma,\xi_\mathrm{cut},\lambda\}=\{0.1\%,8,10^{-4}\}$,
$\{1\%,6,10^{-2}\}$, and $\{5\%,4,0.3\}$. The pointwise relative
errors in Figs.\;\ref{fig:cpe_results}(d) and\;\ref{fig:cpe_results}(f) are practically flat across the six
decades of the $\tau$ window, and lie below the input noise level
$\sigma$ at every $\tau$ tested (dashed horizontal lines), confirming
that the Tikhonov-Hann stabilization acts as a useful low-pass
denoising filter.

\subsection{{DC model DRT and DCT recovery}}

\begin{figure*}[t]
  \centering
    \includegraphics[width=\textwidth]{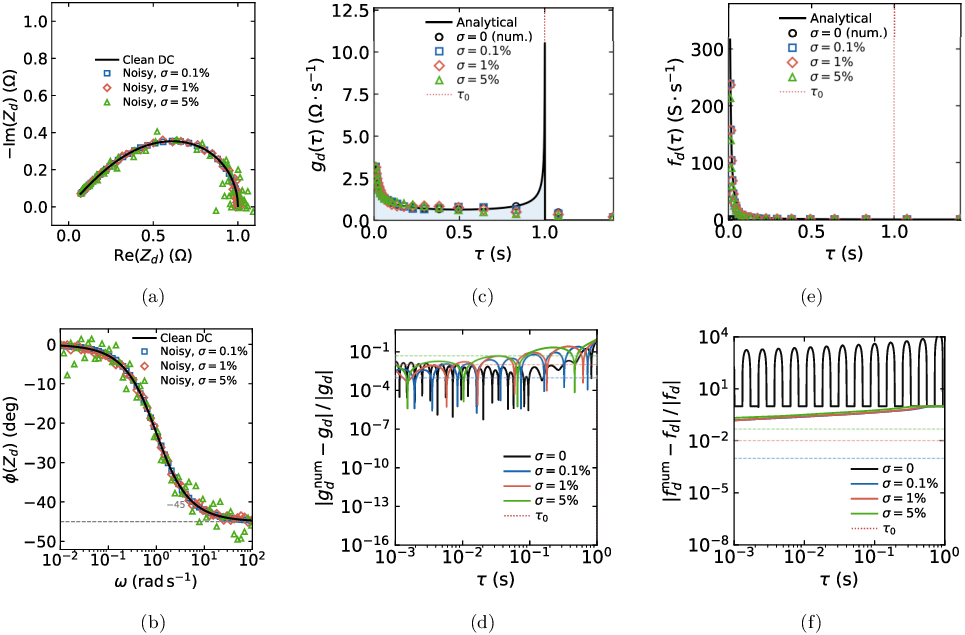}
%  \includegraphics[height=0.285\linewidth]{figures/dc_panel_a_nyquist.pdf}\hfill
%  \includegraphics[height=0.285\linewidth]{figures/dc_panel_c_drt.pdf}\hfill
%  \includegraphics[height=0.285\linewidth]{figures/dc_panel_e_dct.pdf}\\[2pt]
%  \makebox[0.325\linewidth]{\small {(a)}}\hfill
%  \makebox[0.325\linewidth]{\small {(c)}}\hfill
%  \makebox[0.325\linewidth]{\small {(e)}}\\[6pt]
%  \includegraphics[height=0.285\linewidth]{figures/dc_panel_b_bode.pdf}\hfill
%  \includegraphics[height=0.285\linewidth]{figures/dc_panel_d_drt_err.pdf}\hfill
%  \includegraphics[height=0.285\linewidth]{figures/dc_panel_f_dct_err.pdf}\\[2pt]
%  \makebox[0.325\linewidth]{\small {(b)}}\hfill
%  \makebox[0.325\linewidth]{\small {(d)}}\hfill
%  \makebox[0.325\linewidth]{\small {(f)}}
  \caption{Numerical results for the Davidson-Cole model (of impedance $Z_{{d}}(j\omega)= {R_0}{(1+ j \tau_0 \omega)^{-\alpha}}$) with
  $R_0=1\,\Omega$, $\tau_0=1\,\mathrm{s}$, $\alpha=0.5$, and noise
  levels $\sigma\in\{0,0.1\%,1\%,5\%\}$.
  \textbf{(a)}~Nyquist plot of $Z_d(  j \omega_k)$:   clean DC shows a depressed semi-arc
   (solid black), and  open markers show the noisy
  data.
  \textbf{(b)}~corresponding Bode phase vs.\ $\omega$ (log
  scale); the phase is frequency-dependent and approaches
  $-90\alpha=-45^\circ$ (dashed   line) asymptotically at high  frequency.
  \textbf{(c)}~DRT $g_d(\tau)$ recovery on linear axes: solid black
  curve is the analytical Eq.~\eqref{eq:gd};  markers are the   Mellin-FFT recovery including the noise-free case ($\sigma=0$). 
  \textbf{(d)}~Pointwise relative error of the DRT recovery; dashed horizontal lines mark the input noise level $\sigma$.
  The error window is restricted to $\tau<\tau_0$ where the
  distribution is non-zero.
  \textbf{(e)}~DCT $f_d(\tau)$ recovery; same conventions as~(c).
  \textbf{(f)}~Pointwise relative error of the DCT recovery; same
  conventions as~(d); the noise-free error stays near
  $10^{-13}$--$10^{-14}$}
  \label{fig:dc_results}
\end{figure*}

 We now validate the inversion procedure numerically for the DC model.  
% which provides a more demanding test than the CPE because its DRT and DCT are compactly supported distributions with a sharp truncation at $\tau=\tau_0$ and a power-law singularity $(\tau_0/\tau-1)^{-\alpha}$ at that endpoint. Unlike the CPE, the DC DRT has a finite integral equal to $R_0$, and the distribution is concentrated on a bounded interval, making the inversion sensitive to the endpoint singularity.
 We use $R_0=1\,\Omega$, $\tau_0=1\,\mathrm{s}$, $\alpha=0.5$.
The DC impedance is $Z_d(  j \omega)=R_0/(1+  j \tau_0\omega)^\alpha$
and the corresponding admittance is
$Y_d(  j \omega)=R_0^{-1}(1+  j \tau_0\omega)^\alpha$. The analytical
DRT and DCT are given by Eqs.~\eqref{eq:gd} and~\eqref{eq:fd}
respectively. The same multiplicative noise model as in
Eq.~\eqref{eq:noise-Z} is used with
$\sigma\in\{0,0.1\%,1\%,5\%\}$. 
 The DC DRT Mellin transform has fundamental strip
$0<\Re(s)<\alpha=0.5$, so we choose the inversion abscissa at the
centre of the strip, $\sigma_{h,\mathrm{DRT}}=1-\alpha/2=0.75$.
Similarly, for the DC DCT, the contour is placed at
$\sigma_{h,\mathrm{DCT}}=1+\alpha/2=1.25$. 
%Because the DC impedance's Mellin transform is a smooth Gamma-function ratio (not a delta function), the inversion is better conditioned than for the CPE, and only moderate regularisation is required. 
For the noise-free case,
we use $\{\xi_\mathrm{cut},\lambda\}=\{12,10^{-20}\}$; for the noisy
cases, $\{\sigma,\xi_\mathrm{cut},\lambda\}=\{0.1\%,9,10^{-4}\}$,
$\{1\%,7,10^{-2}\}$, $\{5\%,5,0.2\}$.

 The results are shown in Fig.~\ref{fig:dc_results}. Figs.~\ref{fig:dc_results}(a) and ~\ref{fig:dc_results}(b) show the Nyquist plot and Bode phase of the DC impedance for $R_0=1\,\Omega$, $\tau_0=1\,\mathrm{s}$, $\alpha=0.5$, along with noisy data with $\sigma\in\{0.1\%,1\%,5\%\}$. 
 The Nyquist arc is a curved with $Z=R_0$ as $\omega\to 0$ and at $Z=0$ as $\omega\to\infty$, and the
Bode phase is frequency-dependent, approaching $-90\alpha=-45^\circ$ asymptotically. 
Figs.\;\ref{fig:dc_results}(c) and \ref{fig:dc_results}(e) show the DRT and DCT recovery on linear axes. The Mellin-FFT method reproduces the analytical distributions closely up to $\tau\approx\tau_0$, including the steep rise near the truncation point $\tau_0=1\,\mathrm{s}$ (marked by the red dotted line). The noise-free recovery ($\sigma=0$) sits essentially on the analytical curve throughout the interval. At larger noise levels, the recovered distributions deviate primarily near the singularity at $\tau_0$, where the regularization smooths the sharp endpoint. The pointwise relative errors in Figs.~\ref{fig:dc_results}(d) and \ref{fig:dc_results}(f) confirm that the error remains below the input noise level $\sigma$ over most
of the $\tau$ window, and lies near the floating-point limit
($\sim 10^{-13}$) for the noise-free case.

\section{Conclusion}

We have shown that the distribution of relaxation times (DRT), the distribution of capacitive times (DCT), and the distribution of diffusion times (DDT) problems all admit an exact Mellin-domain formulation valid for arbitrary immittance functions. The key observation is that all standard immittance kernels depend on the product $\omega\tau$ rather than on $\omega$ and $\tau$ independently, which is precisely the hallmark of a Mellin convolution. This structure, which has not been previously exploited in the EIS literature, makes the Mellin transform the most natural spectral framework for immittance inversion. The inversion problem is then reduced to algebraic operations in Mellin space: the Mellin transform of the immittance is divided by the Mellin transform of the kernel, followed by an argument shift $s\mapsto 1-s$, and then an inverse Mellin transform that leads to the desired distribution function (see Fig.~\ref{fig:flowchart}). We verified the approach analytically on a few examples of impedance and admittance functions, including the CPE and the DC model for both DRT and DCT recovery, and the finite-length Warburg admittance of a planar electrode with blocking boundary conditions for the DDT problem.
 Numerical validation using the fast Mellin transform via FFT was carried out for both the CPE and the DC model, covering DRT and DCT recovery under clean and noisy conditions ($\sigma$ up to 5\%). For the CPE, the noise-free error reaches the floating-point limit ($\sim10^{-14}$), and for the DC model, the method recovers the distribution faithfully, with errors below the input noise level across most of the timescale window. 
Two practical limitations should be noted: (i) the method requires a sufficiently wide frequency window  to capture the significant Mellin frequency content of the spectrum, and (ii) the inversion abscissa $\sigma_h$ must be chosen to lie within the fundamental strip of the distribution's Mellin transform. 

%\bibliography{bib}


\end{document}